\documentclass[12pt,preprint]{aastex}
\shorttitle{HD~16429~A}
\shortauthors{McSwain}

\begin{document}

\received{}
\accepted{}

\title{The Massive Triple Star System HD~16429~A}

\author{M. V. McSwain\altaffilmark{1}} 

\affil{Center for High Angular Resolution Astronomy\\
Department of Physics and Astronomy, P.O. Box 4106 \\
Georgia State University, Atlanta, GA  30302-4106\\
Electronic mail: mcswain@chara.gsu.edu}

\altaffiltext{1}{Visiting Astronomer, Kitt Peak National Observatory,
National Optical Astronomy Observatories, operated by the Association
of Universities for Research in Astronomy, Inc., under contract with
the National Science Foundation.}

\slugcomment{Submitted to ApJ}
\paperid{}

%%%%%%%%%%%%%%%%%%%%%%%%%%%%%%%%%%%%%%%%%%%%%%%%%%%%%%%%%%%%%%%

\begin{abstract}

HD~16429~A is a triple star system consisting of a single-lined
spectroscopic binary and a widely separated third component, previously
identified via speckle interferometry.  Here I present the first orbital 
elements for the unblended spectroscopic binary as well as estimates of 
the spectral types and relative flux contributions for each visible 
component based upon a Doppler tomographic reconstruction of their 
spectra.  There are several stars around HD~16429~A, including the nearby 
Be X-ray binary and microquasar LS~I~+$61^{\circ}303$, which all probably 
belong to a sub-cluster within the Cas~OB6 association.

\end{abstract}


%%%%%%%%%%%%%%%%%%%%%%%%%%%%%%%%%%%%%%%%%%%%%%%%%%%%%%%%%%%%%%%

\section{Introduction}                              % Section 1

O-type stars are most commonly found within young open clusters and OB 
associations in the spiral arms of the Galaxy \citep{gie87}, and the 
binary frequency among these cluster and association members is 
significantly higher (between $70-75\%$) than among field O stars 
($19-22\%$) and runaways (about $8\%$) \citep{gie87, gar94, mas98}. 
Multiple O star systems in clusters are particularly influential in the 
dynamics of the cluster because stars will be ejected both through 
supernova explosions in close binaries and through close gravitational 
encounters of stars with binaries.  Triple O star systems containing a 
close pair and a more widely separated component are particularly 
important sites of gravitational interactions \citep{mas98}.

HD~16429~A (HIP~12495, V482~Cas, SAO~12383, BD+$60^{\circ}541$) is an 
O9.5 II ((n)) star \citep{wal76} and a member of the Cas OB6 association 
\citep{gar92}.  It has two optical companions, component B at a separation 
of 6.66 arcsec, and C at a separation of 53.2 arcsec \citep{mas98}.  
HD~16429~B is an F4 V star \citep{mei68}, which indicates that it is not 
physically associated with A.  Component C is more likely associated based 
upon its magnitude and color (see \S 4).  The A component was itself found 
to be a speckle binary with a separation of 0.30 arcsec \citep{mas98}.  
\citet{gie86} found that one component of A may also be a spectroscopic 
binary, making it a triple system, but they were unable to identify the 
period of this close pair.  All of the components of the heirarchical 
HD~16429 system are shown in Figure 1.  Other observations have shown 
HD~16429~A to be a radio emitter \citep{mar98} as well as a weak X-ray 
source \citep{gol95}.  The radio and X-ray emission probably originates in 
the stellar winds \citep{mar98}.

\placefigure{fig1}	% Heirarchical diagram of HD 16429 system

Here I present evidence to confirm that HD~16429~A is a triple star 
system with a spectroscopic binary component (HD~16429~Ab1) and a radial 
velocity stationary companion (HD~16429~Aa).  The Aa and Ab1 components 
were isolated using a Doppler tomography algorithm, but the Ab2 component
remains hidden.  Approximate spectral types and relative flux 
contributions are presented for the two visible components.  Finally, I 
discuss the stars in the immediate vicinity of HD~16429~A and their 
relationship to the Cas~OB6 association.

%%%%%%%%%%%%%%%%%%%%%%%%%%%%%%%%%%%%%%%%%%%%%%%%%%%%%%%%%%%%%%%

\section{Observations and Blended Line Radial Velocities}	% Section 2

Thirty-five optical spectra of HD~16429~A were obtained with the Kitt Peak 
National Observatory 0.9-m Coud\'{e} Feed Telescope during runs in 2000 
October and 2000 December.  The spectra have a resolving power $R=\lambda 
/ \delta \lambda = 9500$ using the long collimator, grating B (in second 
order with order sorting filter OG550), camera 5, and the F3KB CCD, a Ford 
Aerospace $3072\times 1024$ device.  These arrangements produced a 
spectral coverage of $6440 - 7105$ \AA.  Usually two exposures of 15 
minutes duration were taken each night, separated by $2-3$ hours.  The 
spectrum of the rapidly rotating A-type star, $\zeta$~Aql, was observed 
each night for removal of atmospheric water vapor and O$_2$ bands.  Also, 
the O-type stars HD~34078 and HD~47839 were observed as spectral 
comparisons.  Each set of observations was accompanied by numerous bias, 
flat field, and Th~Ar comparison lamp calibration frames.  The dates of 
observation are given in Table~1. 

\placetable{tab1}      % Table 1 - Observations

The spectra were extracted and calibrated using standard routines in
IRAF\footnote{IRAF is distributed by the National Optical Astronomy
Observatories, which is operated by the Association of Universities for
Research in Astronomy, Inc., under cooperative agreement with the National
Science Foundation.}.  All the spectra were rectified to a unit continuum
by fitting line-free regions.  The removal of atmospheric lines was done
by creating a library of $\zeta$~Aql spectra from each run, removing the
broad stellar features from these, and then dividing each target spectrum
by the modified atmospheric spectrum that most closely matched the target
spectrum in a selected region dominated by atmospheric absorptions.  
Some glitches appeared as artifacts of the removal of the atmospheric 
telluric lines, and these were excised by linear interpolation.  The 
spectra from each run were then transformed to a common heliocentric 
wavelength grid.  

Radial velocities were measured by fitting a parabola to the line cores 
of H$\alpha$, \ion{He}{1} $\lambda 6678$, and \ion{He}{1} $\lambda
7065$, but several factors made it difficult to obtain consistent
measurements.  The H$\alpha$ line shows variable P Cygni emission that
pushes the line's center blueward.  Due to this certain contamination, 
radial velocities from the H$\alpha$ line were not used.  In addition, 
measurements of the \ion{He}{1} $\lambda 7065$ line have a slightly lower 
mean radial velocity than the mean for the \ion{He}{1} $\lambda 6678$ 
line.  This could be due to different relative line strengths of the 
\ion{He}{1} lines in the blended stars,  contamination from \ion{He}{2} 
$\lambda 6683$ absorption in one component, or a radial velocity gradient 
due to outflowing winds (see \S 4).  However, the difference in their mean 
velocities is small and only corresponds to a difference of 1.4 pixels on 
the chip, so the mean velocity of the two \ion{He}{1} lines was used in 
this analysis.

To account for small deviations in radial velocity due to the wavelength
calibration, the interstellar lines present in the spectra were inspected.
A large number of interstellar features were identified from the lists of 
\citet{her75}, \citet{her91}, \citet{mor91}, and especially \citet{gal00}.  
An interstellar spectrum was created by extracting these lines from the 
mean HD~16429 spectrum, and then it was cross correlated with each 
individual spectrum to measure any systematic radial velocity shift 
(generally $<7$~km~s$^{-1}$).  The velocity shift was subtracted from the 
measured radial velocities, and the resulting velocities are shown in the 
third column of Table 1.  In the next section, I will show that these 
radial velocities actually represent the results of line blending between 
a single-lined spectroscopic binary and a stationary component.

%%%%%%%%%%%%%%%%%%%%%%%%%%%%%%%%%%%%%%%%%%%%%%%%%%%%%%%%%%%%%%%

\section{Orbital Elements from a Two Component Model}	% Section 3

Visual inspection of the spectra revealed radial velocity variations 
indicative of an orbital period of about 3 days, but a more precise period 
search was done using a version of the discrete Fourier transform and 
CLEAN deconvolution algorithm of \citet{rob87} (written in 
IDL\footnote{IDL is a registered trademark of Research Systems, Inc.} by 
A.\ W.\ Fullerton).  The most significant signal found in the CLEANed 
power spectrum occurs at the period $P = 3.051$~d.  Relics of this signal 
also occur at integer multiples of this frequency, but the true period 
cannot be shorter because there are no significant changes in radial 
velocity over the $2-3$ hour intervals between the observations from a 
given night.

This period was used with the non-linear, least-squares fitting program of 
\citet{mor74} to solve for the orbital elements.   The best fit was 
obtained by allowing the period to vary slightly, to an improved value of 
$P = 3.0539$~d, and assigning only half weight to one point with very 
large scatter near 0.95 phase.  The resulting preliminary orbital elements 
are listed in the second column of Table~2, and the radial velocity curve 
is illustrated in Figure~2 with the dotted line.  Note that this curve has 
been shifted in phase so that the epoch, $T$, corresponds to the final 
time of periastron, discussed below.  The observed minus calculated 
velocity errors ($O-C$) based on this preliminary solution are given in 
Table~1, column~4.  The root-mean-square of the velocity residuals is 
comparable to the scatter in the best measured lines.  

\placetable{tab2}      % Table 2 - Orbital Elements

\placefigure{fig2}     % Figure 2 - Radial velocity curve of combined system

There are a number of other radial velocity measurements for HD~16429~A 
in the literature (10 by \citet{abt72}, 15 by \citet{gie86}, and 1 by 
\citet{con77}), so these measurements were combined with those in Table~1 
to refine the orbital period.  Since each investigation used different 
spectral lines and wavelength regions to measure velocities, the results 
may have systematic differences.  However, the agreement between data sets 
was reasonable, and including the additional measurements significantly 
improved the orbital period, $P = 3.05442~\pm~0.00005$~d.

The speckle results \citep{mas98} indicate that at least two bright 
components were recorded in these spectra, so I created grayscale images 
of the spectral lines as a function of orbital phase to search for 
evidence of multiple components.  Figure~3 shows the plot of the 
\ion{He}{1} $\lambda 6678$ line.  It clearly reveals signs of a 
single-lined spectroscopic binary, Ab1, superimposed upon a stationary 
component, Aa, with similar systemic velocity, consistent with 
expectations from the speckle observations.  Because the stationary 
component's lines are blended with the moving component, the preliminary 
radial velocity semiamplitude of the moving component is probably greatly 
underestimated, and perhaps the preliminary eccentricity and longitude of 
periastron are misrepresented as well.  

\placefigure{fig3}     % Figure 3 - Grayscale plot of blend

To determine the true orbital elements of the Ab1 component, the Aa 
component was isolated using a two-component tomographic separation and 
then subtracted from the observations.  Tomographic separation relies upon 
the radial velocity estimates of the stars and their flux ratio to 
recreate the spectra of the individual stars \citep{bag91}.  The algorithm 
uses the observed composite spectra, approximate flux ratios of the stars, 
the center of mass velocity, $V_{0}$, and the velocity semi-amplitude, 
$K$, of component Ab1 to calculate the solution using an iterative least 
squares technique.  For the Aa component, the velocity amplitude was set 
to 0.  A three-component separation was not attempted because of the 
extreme faintness of the Ab2 spectroscopic component.  Even the 
two-component separation was difficult because the true velocity 
amplitude, $K$, of Ab1 is unknown, and visual inspection of Figure 2 
reveals that it could be as high as 200~km~s$^{-1}$.  The tomographic 
separation is not very sensitive to the choice of the relative flux 
contribution of the components.  Initially, an equal flux contribution was 
assumed, and the value was later refined according to the methods 
discussed in the following section.

I computed tomographic separations for a grid of assumed semi-amplitudes 
for $30 < K < 200$~km~s$^{-1}$.  In order to identify the best fit value 
of $K$, I used the separated components to recreate a set of synthetic 
spectra for each observation which was then compared to the original 
spectra.  The \ion{He}{1} $\lambda 6678$ and \ion{He}{1} $\lambda 7065$ 
line regions were used separately to compute the $\chi^{2}$ error between 
the observed and recreated spectra.  The errors associated with the 
\ion{He}{1} $\lambda 7065$ line were slightly larger because that line is 
located in a region with many atmospheric telluric lines, so it was 
correspondingly noisier after the telluric removal process.  The values of 
$K$ corresponding to the minimum error for each line were then averaged to 
obtain the optimal value of $K~=~124~\pm~25$~km~s$^{-1}$.  

Once the Aa component was isolated, I subtracted its profile from the 
observed spectra to attain profiles of the Ab1 component alone, and then I 
measured the radial velocities of Ab1 in the same way as before.  These 
unblended radial velocities are listed in column 5 of Table~1.  The new 
orbital elements of the spectroscopic binary were calculated using the 
same technique discussed above and are included in the last column of 
Table~2.  Table 1, column 2 lists the orbital phase from periastron based 
on this final solution.  Although the optimal $K$ is 
$124~\pm~25$~km~s$^{-1}$, the orbital fitting program found a slightly 
higher value of $K~=~136~\pm~6$~km~s$^{-1}$, but these estimates are 
consistent within errors.  Figure~2 includes the final radial velocities 
for the moving component as the solid points and the theoretical radial 
velocity curve as the solid line, and the new grayscale plot of the 
isolated moving component is shown in Figure~4.  The signal to noise ratio 
is too low to positively identify the anti-phase motion of the 
spectroscopic companion Ab2 in this plot.

\placefigure{fig4}     % Figure 4 - Grayscale plot of moving component

\placefigure{fig5}	% Figure 5 - Separated spectra, standard spectra

%%%%%%%%%%%%%%%%%%%%%%%%%%%%%%%%%%%%%%%%%%%%%%%%%%%%%%%%%%%%%%%

\section{Discussion}                                % Section 4

The flux ratio of the two visible components of HD~16429~A was determined 
using their observed speckle magnitudes, $\delta m_{V} = 0.5 \pm 0.2$ 
(Mason \& Hartkopf 2001, private communication).  The stationary star 
contributes a stronger signal in the spectra, and it is also the brighter 
component with 61~$\pm~5~\%$ of the overall luminosity.  This star is 
HD~16429~Aa described by \citet{mas98}, whereas the visible moving 
component is HD~16429~Ab1.  The flux ratio can also be estimated using the 
equivalent width of each component's \ion{He}{1} $\lambda 6678$ line.  
\citet{con74} found that in late O stars of all luminosity classes, the 
equivalent width of this line is roughly constant (but with a large 
scatter), and I adopted the method of \citet{pet39} to determine the flux 
ratio of the stars from their observed equivalent widths.  Assuming an 
intensity ratio $r~=~F_{2}/F_{1}$ for the tomographic reconstruction, the 
resulting equivalent widths of the reconstructed moving and stationary 
component spectra are $A_{1}$ and $A_{2}$, respectively.  The relative 
continuum intensities are then $1/(1+r)$ and $r/(1+r)$, respectively, and 
the absolute line strengths are $s_{1}~=~A_{1}/(1+r)$ and $s_{2}~=~
rA_{2}/(1+r)$.  Because the true strengths of the lines are the same in 
each star, the true intensity ratio is $r_{1}~=~rA_{2}/A_{1}$.  By 
integrating over the \ion{He}{1} $\lambda 6678$ line in each component and 
comparing their equivalent widths, the contribution of the stationary 
component was found to be about $70~\pm~7~\%$ of the system's total 
luminosity.  This value agrees reasonably well with that obtained from 
speckle observations, but is less accurate because of the intrinsic 
scatter in the \ion{He}{1} $\lambda 6678$ widths, so I adopted the 
flux ratio from the speckle observations.

Figure~5 presents a plot of the reconstructed spectra, using the final $K 
= 136$~km~s$^{-1}$ and normalized for their flux contribution.  The 
figure also shows the Aa and Ab1 components compared to two stars of 
similar spectral type, HD 34078 and HD 47839.  The comparison spectra have 
been smoothed using a Gaussian profile so that their FWHM and line depths 
more closely match those of the separated components.  This smoothing does 
not significantly affect the equivalent widths of the lines.  The spectra 
of the Aa and Ab1 components have also been smoothed slightly to reduce 
noise.

The tomographic separation of the spectra revealed that both visible 
components of HD~16429~A have strong H$\alpha$ and \ion {He}{1} lines.  
The smoothed spectra of HD~34078 and HD~47839 have \ion {He}{1} lines of 
similar strength, although stronger H$\alpha$ profiles.  H$\alpha$ 
emission is common in luminous O stars (luminosity classes I, II, and 
III), so the difference in line strengths is likely due to emission in 
HD~16429~A.  (HD~34078 and HD~47839 are both main sequence O stars with 
no emission.)  The emission may not have been distributed properly 
between the two component stars during the tomographic separation because 
it varies on a non-orbital timescale.  The original blended spectra showed 
striking P~Cygni emission in the H$\alpha$ line, which is associated 
almost entirely with the Aa component in the reconstructions.  This 
widely-separated member of the triple system is probably the more evolved 
of the group since it has such strong emission from stellar winds.  Since 
it is the more luminous, the stationary component corresponds to the O9.5 
II type that was assigned to the system by \citet{wal76}.  The ((n)) 
suffix used by Walborn to indicate slightly broadened lines probably 
reflects the line blending with the Ab1 component.  The \ion {He}{1} line 
strengths are similar to those in HD 34078, an O9.5 V star.  The 
absence of significant \ion {He}{2} provides further confirmation of the 
spectral type.

The separation also revealed weak \ion {He}{2} $\lambda\lambda 6527$, 
6683, and 6891 absorption in the Ab1 component.  \ion {He}{2} lines only 
appear in O spectra, disappearing entirely in the earliest B spectra.  
Their presence indicates that this star is hotter than the brighter Aa 
star.  Comparing the strengths of these lines with those in HD 47839 
reveals that the moving component is a slightly later, O8 spectral type, 
although the luminosity class cannot be specified by the comparison.  
However, the difference in speckle magnitudes is consistent with an 
O8~III-V classification for HD~16429~Ab1 \citep{how89}.  

The second spectroscopic companion, Ab2, remains hidden even after
the tomographic separation (Fig.~4), but the statistical method of
\citet{maz92} suggests that it is a massive, early B-type star.
For single-lined spectroscopic binaries consisting of a primary of mass 
$M_{1}$ and a secondary of mass $M_{2}$, with a mass ratio 
$q = M_{2}/M_{1}$, the mass function is given by $f(M_{2})~=~ 
M_{1}$~sin~$^{3}~i$~$q^{3}/(1~+~q)^{2}$.  It cannot be solved uniquely for 
both $q$ and $i$, but Mazeh \& Goldberg recommend an iterative solution 
based on a random distribution of both parameters.  Given the known mass 
function and the primary mass, the mass ratio can be calculated for each 
possible inclination and then averaged over the range of inclinations.  
Using a mass of $28-33 M_\sun$ for the O8~III-V primary \citep{how89} and 
$f(M_{2}) = 0.76~\pm~0.11$ from the final orbital solution, the probable 
mass ratio is 0.6.  This is inconsistent with a low mass compact 
companion, but suggests instead a $17-20 M_\sun$ early B-type star 
\citep{how89}.  The Ab2 component remains hidden because its flux is 
overwhelmed by the two brighter O stars.  If these masses are adopted, 
then the orbital inclination is approximately $i~=~42^\circ$, and no 
eclipses are expected (or observed by Hipparcos; \citealt{per97}).

There are several stars in the vicinity of HD~16429~A, including the Be 
X-ray binary LS~I~+$61^{\circ}303$, that have similar radial velocities, 
proper motions, apparent $V$ magnitudes, and $B-V$ colors (summarized in 
Table~3).  Parabolic fits of the \ion{He}{1} $\lambda 6678$ and 
\ion{He}{1} $\lambda 7065$ lines in the tomographically isolated spectrum 
of the Aa component provided its average radial velocity of 
$-26~\pm~5$~km~s$^{-1}$.  This is somewhat different than the systemic 
velocity of the spectroscopic pair, $-49.5~\pm~3.3$~km~s$^{-1}$.  However, 
the observed difference in $V_0$ in HD~16429~Aa and Ab1 is probably due to 
differing wind velocities of the two stars.  Radial velocities of 
photospheric lines are known to reflect the outward acceleration of the 
line formation region \citep{con77b, boh78}.  LS~I~+$61^{\circ}303$ has a 
systemic velocity $V_{0}~=~-55~\pm~6$~km~s$^{-1}$ \citep{hut81}.  No 
radial velocities are available for HD~16429~C or ALS~7378.  The proper 
motion of each star is from the Simbad database.  For HD~16429~Aa and Ab1, 
$V$ and $B-V$ were found using the values from \citet{hil56} and the flux 
ratio determined above.  For LS~I~+$61^{\circ}303$, those values were 
found by averaging those by \citet{dri75} and \citet{coy83}.  
LS~I~+$61^{\circ}303$ has a somewhat redder $B-V$ than the other stars, 
but it has previously been identified with the Cas OB6 association by 
\citet{ste98}.  $V$ and $B-V$ for the remaining stars are from the Simbad 
database.  Based on these similarities, HD~16429~A and its neighbors 
probably make up a small sub-cluster within the Cas OB6 association.  

The stars in this sub-cluster were probably born in the same generation of 
star formation and therefore have approximately the same age and distance.  
\citet{gar92} used main sequence fitting to determine a distance of 
$2.40~\pm~0.18$~kpc to the Cas OB6 association.  However, \citet{fra91} 
used a kinematical model of the Galaxy to determine the distance to 
LS~I~+$61^{\circ}303$ based on the velocities of \ion{H}{1} absorption 
lines in its spectrum, and they placed that system at a closer distance of 
$2.0~\pm~0.2$~kpc (confirmed by \cite{ste98}).  Therefore I adopted a 
distance of 2.0~kpc for the group.  Placing the stars on evolutionary 
tracks in a color-magnitude diagram is difficult since the reddening is 
variable across the association \citep{han93}.  For HD~16429~A, the 
spectral type of each component was used to obtain its intrinsic $B-V$ 
\citep{weg94}, which was then combined with the observed $B-V$ for the 
system \citep{hil56} to determine its reddening, $E(B-V)$, and absolute 
magnitude, $M_{V}$.  \citet{how83} provides $E(B-V)~=~0.75$ for 
LS~I~+$61^{\circ}303$.  The reddening of HD~16429~C and ALS~7378 was 
estimated assuming that each is a B0 V star (based on their $B$ and $V$ 
magnitudes relative to HD~16429~A) and following the same procedure used 
for HD~16429~A.  Table 3 also lists the $E(B-V)$ and $M_{V}$ for each star 
in the sub-cluster.

\citet{boh81} provides temperatures of massive stars based on their 
spectral types, and I used her calibration to find their temperatures and 
bolometric corrections \citep{how89}.  Their luminosities were found using 
the expression log~$L/L_\sun~=~-0.4(M_{bol}~-~4.74)$.  The evolutionary 
tracks for non-rotating stars of \citet{sch92}, based on log $T$ and log 
$L/L_\sun$, provide estimates of the current and initial stellar masses as 
well as stellar ages, and these are also included in Table 3.  It is 
interesting to note that the initial mass of HD~16429~Aa places a lower 
limit on the initial mass of the neutron star progenitor in 
LS~I~+$61^{\circ}303$.  It must have originated in a star with more than 
39 $M_\sun$ when the group formed $2-4$ Myr ago to have evolved more 
quickly than HD~16429~Aa.

\placetable{tab4}      % Table 3 - Evolutionary Tracks

%%%%%%%%%%%%%%%%%%%%%%%%%%%%%%%%%%%%%%%%%%%%%%%%%%%%%%%%%%%%%%%

\acknowledgments

I especially thank Douglas Gies for his advice and comments regarding this
manuscript.  I also thank David Wingert and Wenjin Huang for taking many 
of the observations, and the KPNO staff for their assistance and travel 
support.  I am also grateful to Dr.\ Alex Fullerton for sharing his period 
search code, and to Brian Mason and William Hartkopf of the U.S. Naval 
Observatory for sending me information from their speckle interferometric 
observations.   This research has made use of the SIMBAD database, 
operated at CDS, Strasbourg, France.  Financial support was provided 
by the National Science Foundation through grant AST$-$0205297 (DRG).  
Institutional support has been provided from the GSU College of Arts and 
Sciences and from the Research Program Enhancement fund of the Board of 
Regents of the University System of Georgia, administered through the GSU 
Office of the Vice President for Research.  I gratefully acknowledge all 
this support.

%%%%%%%%%%%%%%%%%%%%%%%%%%%%%%%%%%%%%%%%%%%%%%%%%%%%%%%%%%%%%%%

% References

%%%%%%%%%%%%%%%%%%%%%%%%%%%%%%%%%%%%%%%%%%%%%%%%%%%%%%%%%%%%%%%

% Tables

\clearpage

% Table 1 - Journal of Spectroscopy
% dates of observations

\begin{deluxetable}{ccrrrr}
\tablewidth{0pc}
\tablecaption{Journal of Spectroscopy \label{tab1}}
\tablehead{
\colhead{Date} &
\colhead{Orbital} &
\colhead{$V_{r}$ (blend)} &
\colhead{$O-C$} & 
\colhead{$V_{r}$ (final)} & 
\colhead{$O-C$} \\
\colhead{(HJD-2,450,000)} &
\colhead{Phase} &
\colhead{(km s$^{-1}$)} &
\colhead{(km s$^{-1}$)} &
\colhead{(km s$^{-1}$)} &
\colhead{(km s$^{-1}$)}
}
\startdata
  1817.877 &  0.334 & \phn         $ -21.4$ &\phn\phs $   3.3$	& \phn         $ -12.5$ &         $ -15.9$ \\
  1818.881 &  0.663 & \phn\phn     $  -4.6$ &\phs     $  14.9$	& \phn\phs     $  16.7$ &         $ -25.7$ \\
  1819.766 &  0.953 & \phn         $ -42.0$ &\phs     $  30.0$	&              $-193.8$ &         $ -11.7$ \\
  1820.879 &  0.317 & \phn         $ -23.9$ &\phn\phs $   3.2$	& \phn         $ -35.4$ &         $ -28.5$ \\
  1821.833 &  0.629 & \phn\phs     $  12.0$ &\phs     $  29.1$	& \phn\phs     $  44.6$ &\phn     $  -8.4$ \\
  1822.857 &  0.964 & \phn         $ -64.7$ &\phn\phs $   9.9$	&              $-201.9$ &         $ -12.3$ \\
  1823.806 &  0.275 & \phn         $ -31.6$ &\phn\phs $   2.3$	& \phn         $ -30.8$ &\phn\phs $   4.4$ \\
  1823.915 &  0.311 & \phn         $ -27.7$ &\phn\phs $   0.2$	& \phn         $ -26.7$ &         $ -16.1$ \\
  1824.817 &  0.606 & \phn\phn     $  -9.4$ &\phn\phs $   6.5$	& \phn\phs     $  38.1$ &         $ -19.9$ \\
  1824.942 &  0.647 & \phn\phn     $  -0.6$ &\phs     $  17.7$	& \phn\phs     $  43.6$ &\phn     $  -4.2$ \\
  1830.834 &  0.576 & \phn\phn     $  -1.8$ &\phs     $  13.0$	& \phn\phs     $  31.6$ &         $ -30.6$ \\
  1830.938 &  0.610 & \phn\phn\phs $   1.0$ &\phs     $  17.1$	& \phn\phs     $  39.7$ &         $ -17.6$ \\
  1888.817 &  0.559 & \phn\phn     $  -0.7$ &\phs     $  13.5$	& \phn\phs     $  82.6$ &\phs     $  19.2$ \\
  1889.816 &  0.886 & \phn         $ -56.9$ &\phn     $  -1.1$	&              $-118.6$ &\phn\phs $   6.1$ \\
  1890.721 &  0.183 & \phn         $ -46.8$ &\phn\phs $   8.3$	&              $-122.1$ &         $ -11.0$ \\
  1890.801 &  0.209 & \phn         $ -48.2$ &\phn\phs $   0.1$	&              $-108.0$ &         $ -19.8$ \\
  1892.715 &  0.835 & \phn         $ -59.0$ &         $ -14.9$	& \phn         $ -94.6$ &         $ -19.9$ \\
  1892.867 &  0.885 & \phn         $ -62.1$ &\phn     $  -6.5$	&              $-126.4$ &\phn     $  -2.7$ \\
  1893.740 &  0.171 & \phn         $ -56.6$ &\phn\phs $   1.5$	&              $-110.8$ &\phs     $  10.4$ \\
  1893.884 &  0.218 & \phn         $ -45.5$ &\phn\phs $   0.4$	& \phn         $ -67.0$ &\phs     $  13.3$ \\
  1894.760 &  0.505 & \phn\phn\phs $   6.5$ &\phs     $  20.4$	& \phn\phs     $  87.7$ &\phs     $  26.0$ \\
  1895.715 &  0.818 & \phn         $ -41.7$ &\phn     $  -1.2$	& \phn         $ -51.5$ &\phn\phs $   6.7$ \\
  1895.804 &  0.847 & \phn         $ -47.2$ &\phn     $  -0.6$	& \phn         $ -63.3$ &\phs     $  22.5$ \\
  1896.645 &  0.122 & \phn         $ -69.0$ &\phn\phs $   2.2$	&              $-154.3$ &\phn\phs $   8.0$ \\
  1896.778 &  0.166 & \phn         $ -61.9$ &\phn     $  -2.3$	&              $-121.8$ &\phn\phs $   3.9$ \\
  1897.646 &  0.450 & \phn\phn     $  -2.4$ &\phs     $  12.9$	& \phn\phs     $  69.5$ &\phs     $  17.9$ \\
  1897.778 &  0.493 & \phn         $ -10.9$ &\phn\phs $   3.1$	& \phn\phs     $  70.3$ &\phs     $  10.1$ \\
  1898.650 &  0.779 & \phn         $ -37.7$ &\phn     $  -4.1$	& \phn         $ -34.7$ &\phn     $  -9.8$ \\
  1898.782 &  0.822 & \phn         $ -47.8$ &\phn     $  -6.4$	& \phn         $ -63.6$ &\phn     $  -1.5$ \\
  1899.652 &  0.107 & \phn         $ -57.1$ &\phs     $  17.8$	&              $-161.2$ &\phs     $  12.6$ \\
  1899.784 &  0.150 & \phn         $ -58.9$ &\phn\phs $   5.0$	&              $-136.9$ &\phn\phs $   2.5$ \\
  1900.647 &  0.433 & \phn         $ -11.5$ &\phn\phs $   4.6$	& \phn\phs     $  70.7$ &\phs     $  24.0$ \\
  1900.778 &  0.476 & \phn\phn     $  -0.6$ &\phs     $  13.8$	& \phn\phs     $  80.4$ &\phs     $  23.0$ \\
  1901.631 &  0.755 & \phn         $ -25.1$ &\phn\phs $   4.7$	& \phn\phs     $  18.2$ &\phs     $  25.1$ \\
  1901.764 &  0.798 & \phn         $ -36.8$ &\phn\phs $   0.1$	& \phn         $ -20.7$ &\phs     $  20.2$ \\
\enddata
\end{deluxetable}

% Table 2 - Orbital Elements
\clearpage

\begin{deluxetable}{lcc}
\tablewidth{0pc}
\tablecaption{Orbital Elements \label{tab2}}
\tablehead{
\colhead{Element} &
\colhead{Blend} &
\colhead{Final Value}
}
\startdata
$P$ (d)                      & 3.054 (5)   	& 3.05442$^a$ 	\\
$T$ (HJD -- 2,450,000)       & 1892.44 (29)	& 1893.22 (10)	\\
$K$ (km s$^{-1}$)            & 35.1 (18) 	& 136 (6) 	\\
$V_0$ (km s$^{-1}$)          & $-33.5$ (11)	& $-50$ (3)	\\
$e$                          & 0.08 (6)		& 0.17 (4)	\\
$\omega$ ($^\circ$)          & 80 (30) 		& 169 (12)	\\
r.m.s. (km s$^{-1}$)         & 6.5 		& 18.0		\\
$f(m)$ ($M_\sun$)            & 0.014 (2)	& 0.76 (11)	\\
$a_1 \sin i$ ($R_\sun$)      & 2.1 (1)	 	& 8.1 (4) 	\\
\enddata
\tablecomments{
The parenthetic numbers are the standard errors in the last digit quoted 
(decimals omitted).  \\
\\
$^a$Fixed.
}
\end{deluxetable}

% Table 3 - Associated Stars Near HD 16429
\clearpage

\begin{deluxetable}{lccccc}
\rotate
\tablewidth{0pc}
\tablecaption{Associated Stars Near HD 16429 \label{tab4}}
\tablehead{
\colhead{Element} &
\colhead{HD 16429 Aa} &
\colhead{HD 16429 Ab1} &
\colhead{HD 16429 C} &
\colhead{ALS 7378} &
\colhead{LS I~$+61^{\circ}~303$}
}
\startdata
Separation (arcmin)		& \nodata	& \nodata	& 0.9		& 1.4		& 3.6		\\
Radial Velocity (km~s$^{-1}$)	& $-26$ (5)	& $-50$ (3)	& \nodata	& $-55$ (6)	& \nodata	\\
$\mu_{\alpha}$ (mas~yr$^{-1}$)	& $-5.8$ (21)	& \nodata	& $-4$ (4)	& 4 (4)		& 0.6 (22)	\\
$\mu_{\delta}$ (mas~yr$^{-1}$)	& 3.1 (12)	& \nodata	& 6.3 (16)	& $-1.7$ (26)	& 1.6 (14)	\\
$V$				& 8.2		& 8.7		& 10.8		& 11.0		& 10.76		\\
$B-V$				& 0.64		& 0.59		& 0.46		& 0.4		& 0.83		\\
$E(B-V)$			& 0.88 (3)	& 0.88 (2)	& 0.72 (2)	& 0.66 (2)	& 0.75 (2)	\\
$M_{V}$ ($d = 2.0$~kpc)		& $-6.0$ (5)	& $-5.5$ (5)	& $-2.9$ (5)	& $-2.6$ (5)	& $-3.1$ (5)	\\
log $T$ (K)			& 4.51 (2)	& 4.55 (1)	& 4.47 (1)	& 4.47 (1)	& 4.47 (1)	\\
log $L/L_\sun$			& 5.57 (26)	& 5.44 (23)	& 4.20 (23)	& 4.08 (23)	& 4.28 (23)	\\
Initial $M$ ($M_\sun$)		& 39 (13)	& 36 (9)	& 13.6 (17)	& 13.0 (16)	& 14.1 (18)	\\
Current $M$ ($M_\sun$)		& 37 (11)	& 35 (8)	& 13.6 (17)	& 13.0 (16)	& 14.1 (18)	\\
Age (Myr)			& 3.8 (9)	& 3.4 (5)	& 1.2 (2.8)	& 2.1 (10)	& 3.0 (18)	\\
\enddata
\tablecomments{
The errors are presented in the same format as Table 2.  Errors in 
$E(B-V)$ and $M_{V}$ are based on errors in the intrinsic $(B-V)$ color 
and distance.  The error in log $T$ is based on the temperature 
difference between adjacent spectral types \citep{boh81}, and the 
remaining errors follow from these.
}
\end{deluxetable}

%%%%%%%%%%%%%%%%%%%%%%%%%%%%%%%%%%%%%%%%%%%%%%%%%%%%%%%%%%%%%%%

% Figure Captions

\clearpage

% Figure 1
% heirarchical system
\begin{figure}
%\plotone{fig1.eps}
\caption{
The various components of the HD~16429 system are shown.  The B component 
is a chance optical alignment and is not physically associated with the 
group.  Its $V$ magnitude is from the Washington Double Star Catalog 
(http://ad.usno.navy.mil/wds/), and all other spectral types and 
magnitudes are discussed in the text.
}
\label{fig1}
\end{figure}

% Figure 2
% radial velocity curve
\begin{figure}
%\plotone{fig2.ps}
\caption{
The radial velocity measurements of the blended components ({\it open
circles}) and the preliminary orbital solution ({\it dotted line}) are
plotted against the final orbital phase.  Also plotted are the radial 
velocity measurements of the isolated spectroscopic component ({\it filled
circles}) and its orbital solution ({\it solid line}).  Phase zero
corresponds to periastron.} 
\label{fig2} 
\end{figure}

% Figure 3
% grayplot of mixed components, He I 6678 line
\begin{figure}
%\plotone{fig3.ps}
\caption{
The \ion {He}{1} $\lambda 6678$ line is shown in linear plots (above) 
and as a grayscale image (below) which shows a blend of the moving and 
stationary components.  The intensity at each velocity in the grayscale 
image is assigned one of 16 gray levels based on its value between the 
minimum (dark) and maximum (bright) observed values.  The intensity  
between observed spectra is calculated by a linear interpolation between 
the closest observed phases (shown by arrows along the right axis).  The 
solid white line shows the theoretical velocity curve of the blended 
solution. 
} 
\label{fig3} \end{figure}

% Figure 4
% grayplot of isolated moving component, He I 6678 line
\begin{figure}
%\plotone{fig4.ps}
\caption{
The \ion {He}{1} $\lambda 6678$ line for the isolated moving component is
shown in the same format as Fig. 3. 
}
\label{fig4}
\end{figure}

% Figure 5
% Separated spectra & spectral standards
\begin{figure}
%\plotone{fig5.ps}
\caption{
The separated Aa and Ab1 components of HD~16429 are shown with the 
spectral comparison stars HD~34078 and HD~47839.  Their rectified fluxes 
are offset vertically for clarity.
}
\label{fig5}
\end{figure}

%%%%%%%%%%%%%%%%%%%%%%%%%%%%%%%%%%%%%%%%%%%%%%%%%%%%%%%%%%%%%%%
% Figures

\clearpage

\setcounter{figure}{0}

\begin{figure}
\plotone{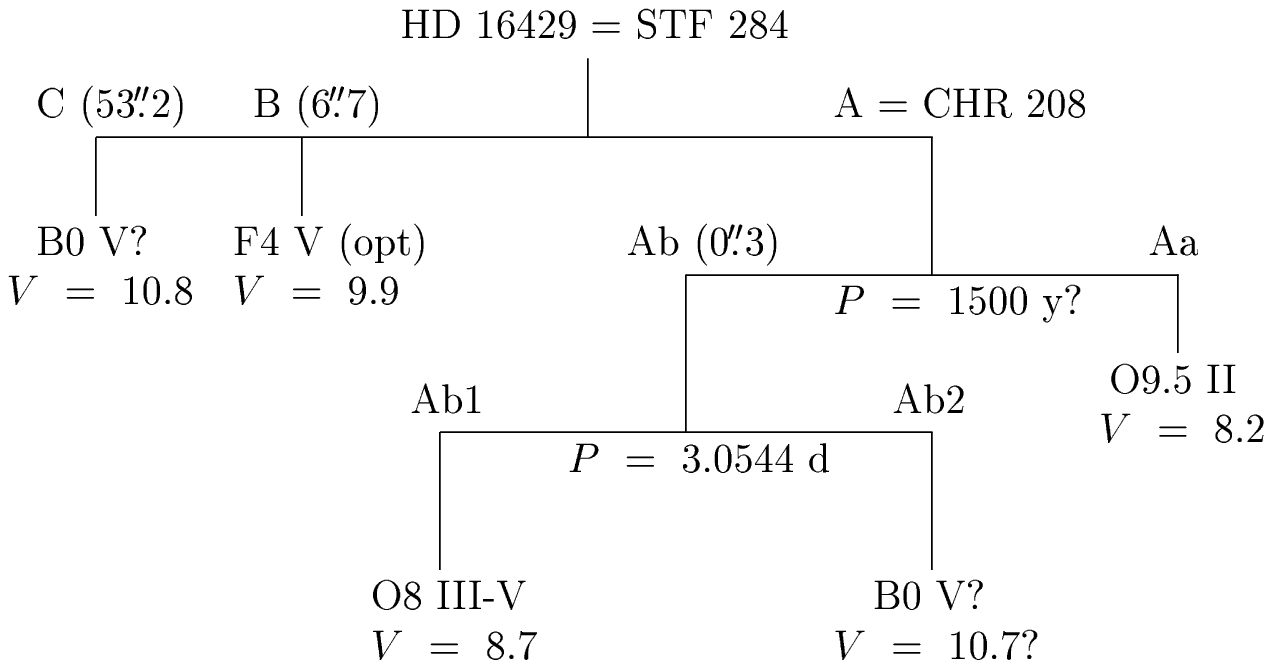}
\caption{}
\end{figure}

\begin{figure}
\plotone{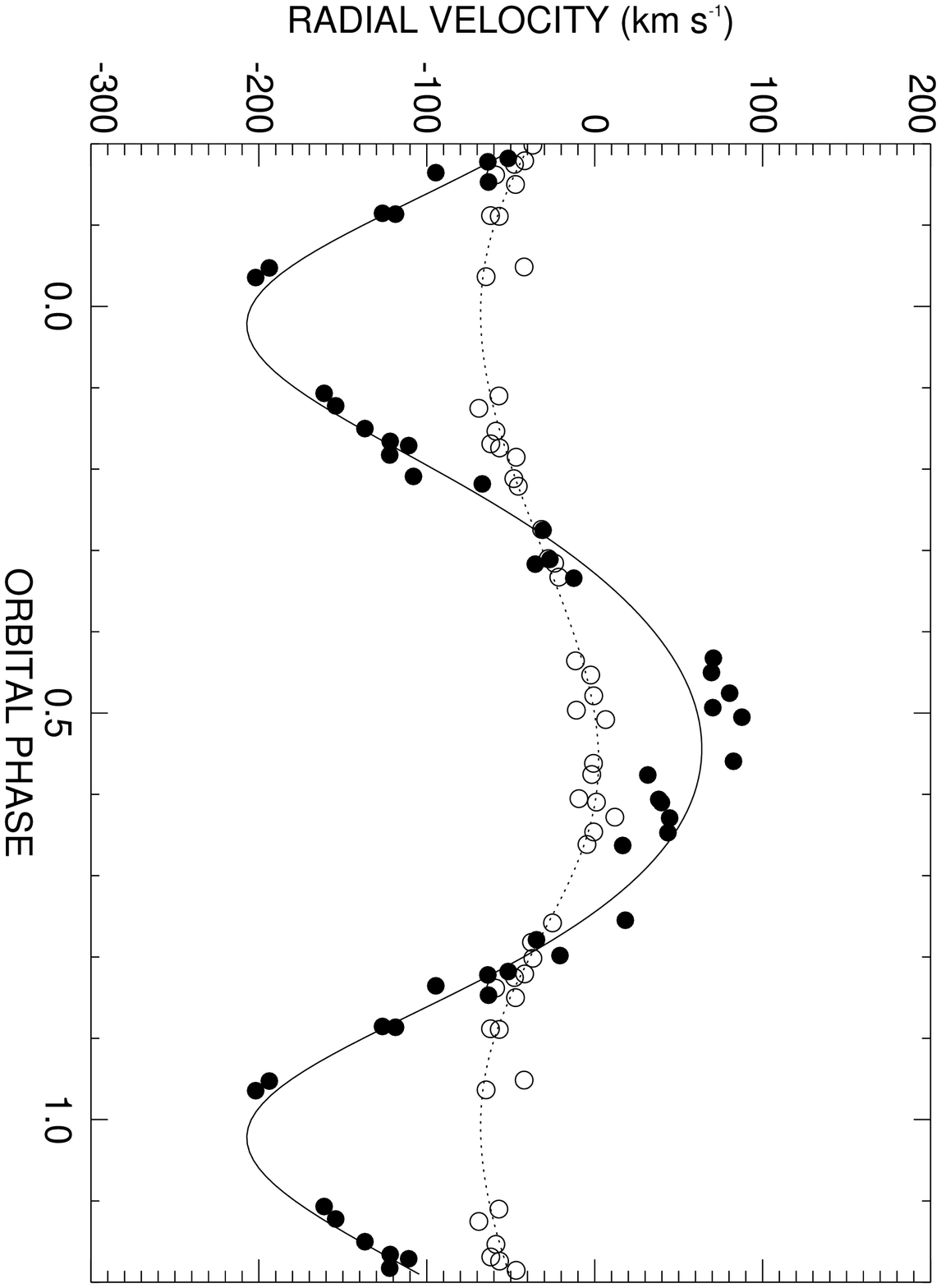}
\caption{}
\end{figure}

\begin{figure}
\plotone{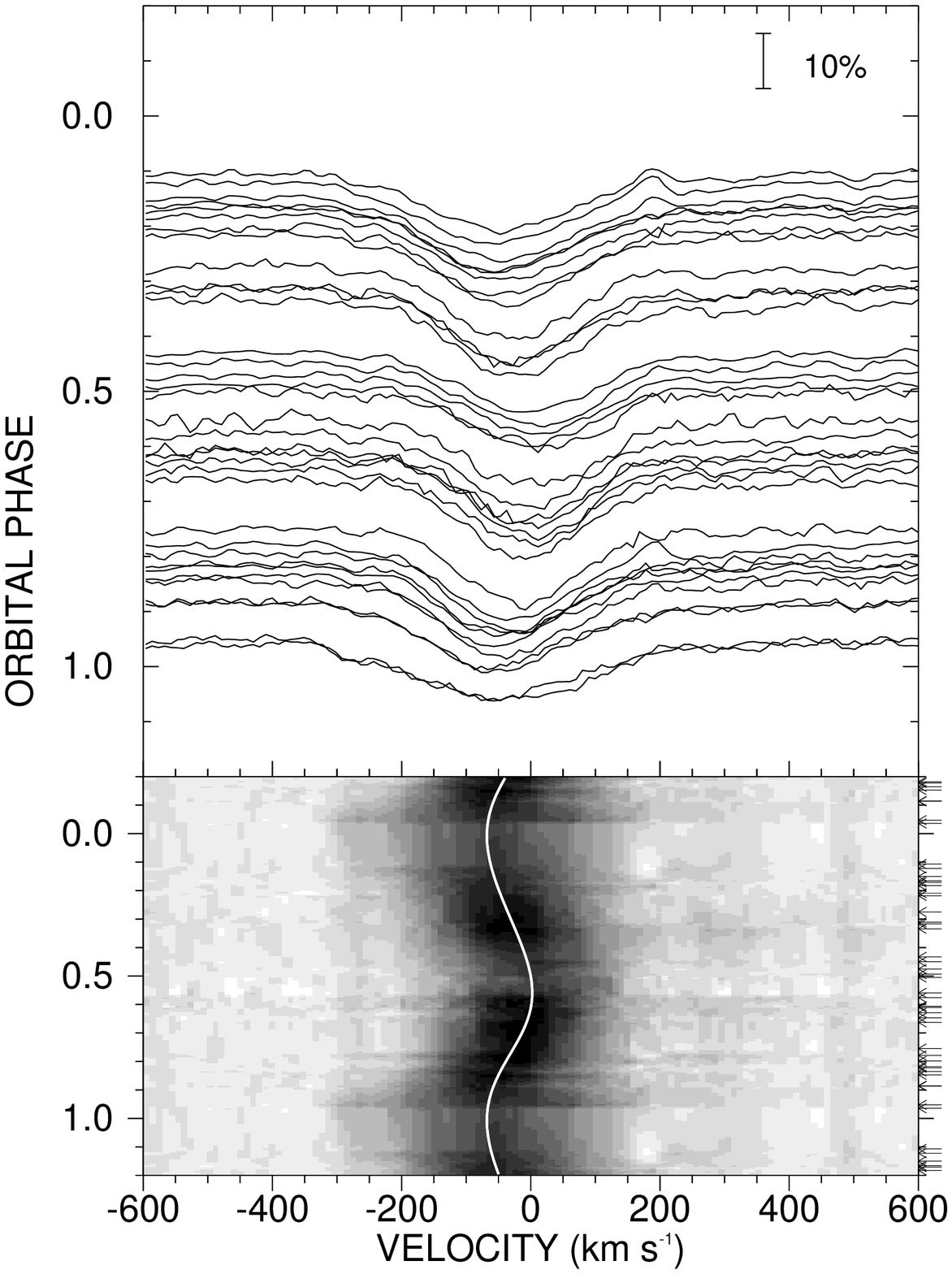}
\caption{}
\end{figure}

\begin{figure}
\plotone{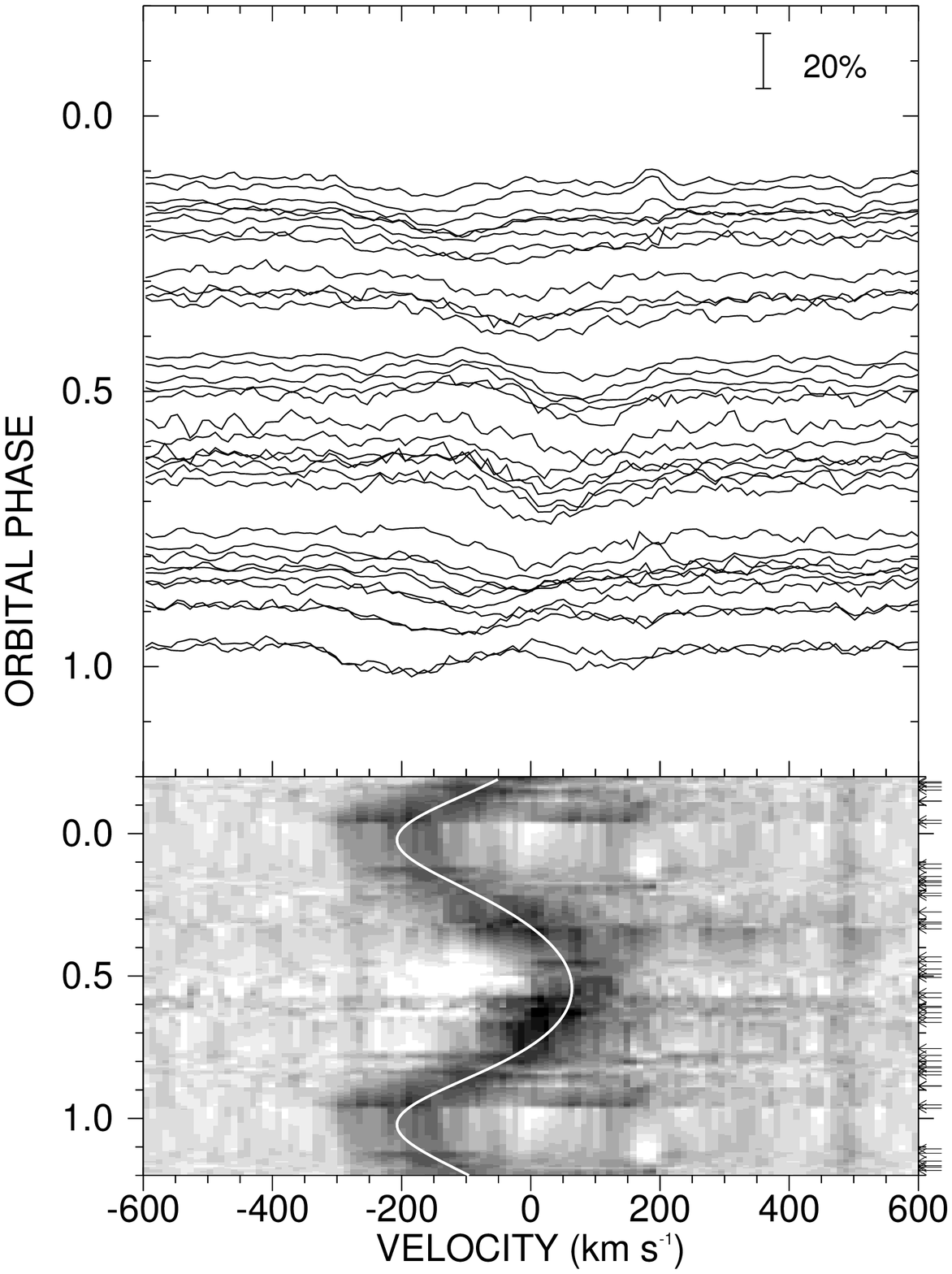}
\caption{}
\end{figure}

\begin{figure}
\plotone{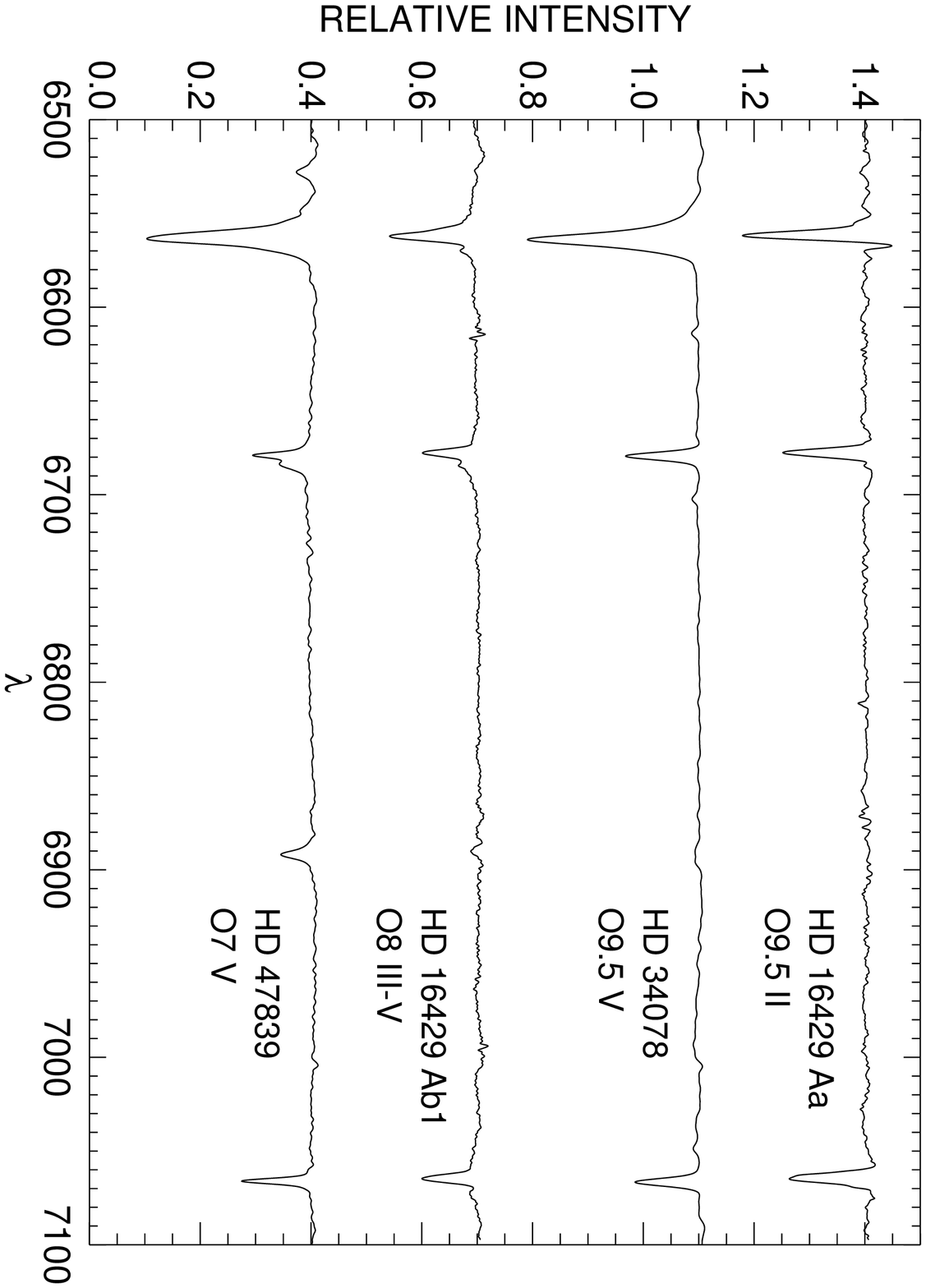}
\caption{}
\end{figure}

%%%%%%%%%%%%%%%%%%%%%%%%%%%%%%%%%%%%%%%%%%%%%%%%%%%%%%%%%%%%%%%

\end{document}